\documentclass[11pt]{article}

\usepackage{jheppub}
\usepackage{amssymb,latexsym}
\usepackage{graphicx}
\usepackage{dcolumn}
\usepackage{bm}

\def \be{\begin{equation}}
\def \ee{\end{equation}}
\def \bea{\begin{eqnarray}}
\def \eea{\end{eqnarray}}
\def \ba{\begin{align}}
\def \ea{\end{align}}
\def \nn{\nonumber}
\def \ie{{\it i.e.}}
\def \eg{{\it e.g.}}
\def \d{{\rm d}}
\def \del{\partial}
\def \na{\nabla}

\begin{document}

\title{Massive Gravity from Higher Derivative Gravity with Boundary Conditions}

\author{Minjoon Park and Lorenzo Sorbo}
\emailAdd{minjoonp@physics.umass.edu, sorbo@physics.umass.edu}
\affiliation{Department of Physics, University of Massachusetts, Amherst, MA 01003}

\date{\today}

\abstract{With an appropriate choice of parameters, a higher derivative theory of gravity can describe a normal massive sector and a ghost massless sector. We show that, when defined on an asymptotically de Sitter spacetime with Dirichlet boundary conditions, such a higher derivative gravity can provide a framework for a unitary theory of massive gravity in four spacetime dimensions. The resulting theory is free not only of higher derivative ghosts but also of the Boulware-Deser mode.}

\maketitle

\section{Introduction}
The seemingly simple task of giving mass to a graviton or a spin-2 field has proved to be rather challenging. To obtain a linearized theory of massive gravity, we add a mass term to the linearized Einstein-Hilbert action. Out of the ten degrees of freedom (DOFs) of the 4D metric perturbations, four are removed by requiring covariant conservation of energy and momentum. But one of the remaining six DOFs being a ghost, we are forced to choose a special form of the mass term, to get the so-called Pauli-Fierz (PF) theory~\cite{Fierz:1939ix} where an extra on-shell symmetry kills the sixth mode. Then, 
\begin{itemize}
\item A massive graviton has more DOFs than the massless one of general relativity (GR), and at the linearized level its coupling to matter is finitely different from that of GR even for arbitrarily small graviton mass. This discrepancy, dubbed as van Dam-Veltman-Zakharov (vDVZ) discontinuity, leads to conclude that massive gravity would be experimentally disfavored~\cite{vanDam:1970vg}.

\item One of the five DOFs of the PF theory can get strongly coupled at too low energy scales~\cite{ArkaniHamed:2002sp}. While this affects the predictivity of the theory, it also signifies that the vDVZ discontinuity -- that was derived assuming the linearized theory is valid at short scales --  should not worry us too much.

\item Upon turning on generic higher order interactions, the extra symmetry of the PF theory gets broken, and the sixth mode or the Boulware-Deser (BD) ghost reappears~\cite{Boulware:1973my}. 

\item The structure or tuning of the Pauli-Fierz mass term is not radiatively stable, and loop corrections from generic quantum interactions will destroy it~\cite{Park:2010rp}.

\item It is highly non-trivial to construct a consistent non-linear completion of the Pauli-Fierz theory.

\end{itemize}

The study of massive gravity is not confined to the somewhat academic challenges listed above:  interest in this field has been resurrected by the advent of  theories of modified gravity as an alternative way of tackling the cosmological constant problem -- in particular, some of the popular higher dimensional modified gravity models are, from the effective 4D theory point of view, PF theories. Therefore the construction of a theoretically and phenomenologically consistent model of massive gravity could affect the discussion about the accelerated expansion of the Universe.

Recently there have been two major developments in the quest for a consistent theory of massive gravity. First, ref.~\cite{deRham:2010kj} formulated a non-linear completion of massive gravity without the BD ghost by introducing an auxiliary metric and an elaborately designed interaction potential between the physical metric and the auxiliary one. The second~\cite{Bergshoeff:2009hq} employed higher derivative gravity (HDG) as a framework of a consistent massive gravity in 3D: HDG once waxed due to its capability of addressing renormalizable quantum gravity~\cite{Stelle:1976gc}, but then waned because it generically contains extra ghost DOFs that destroy the unitarity of the theory. But there being no propagating massless gravitational DOFs in 3D, HDG in 3D has only the massive sector, which can be chosen to be non-ghost-like.

Extending the scheme of \cite{Bergshoeff:2009hq} to 4D is not trivial, because in 4D the massless sector still has two DOFs; if we choose one of the sectors to be normal, the other is bound to contain ghosts. Then there came \cite{Maldacena:2011mk}, whose author took the conformal gravity of (Weyl tensor)${}^2$ which, due to its higher derivative nature, has both a massless and a massive tensor with opposite sign kinetic terms, and put it in a maximally symmetric spacetime. By imposing appropriate boundary conditions to truncate massive ghost modes, \cite{Maldacena:2011mk} showed that conformal gravity can reproduce GR at the classical level. One can expect that combining the ideas of \cite{Bergshoeff:2009hq} and \cite{Maldacena:2011mk} might provide a scheme to obtain a unitary theory of massive gravity in 4D.\footnote{This path is explored in \cite{Deser:2012qg} to construct a theory of a partially massless spin-2 field in a conformal gravity setting.} In the next section we will show that this scheme does indeed work for HDG on a de Sitter (dS) background. Then in \S\ref{sec:adm}, we check that HDG can be free of one of the most notorious problems of massive gravity, the BD mode. We briefly discuss more of the possible issues in \S\ref{sec:conc}.

\section{Higher derivative ghosts}%

In order to make sense of a theory defined with higher time derivatives, we need a way to project out the unwanted ghost DOFs. We will show that for a theory defined on a dS background, imposing Dirichlet boundary conditions on the dS boundary provides such a projection mechanism.

\subsection{Scalar toy model}

Let us first warm up by considering a higher derivative action for a scalar field $\Psi$, which we write in the form
\be\label{eqn:toyact}
S_\Psi=\frac{1}{2}\int\d^4x\sqrt{-g}\Big\{(\Box \Psi)^2 + m^2\del_\mu \Psi\,\na^\mu \Psi  \Big\}\,,
\ee
in a dS background, $\d s^2=\frac{1}{(-H\tau)^2}(-\d\tau^2+\d\vec x\,{}^2)$. By introducing 
\be
\phi\equiv-\frac{\Box\Psi}{m^2}\,,\quad \psi\equiv \Psi+\phi\,,
\ee
we can rewrite it as
\be\label{eqn:act3}
S_\Psi=\frac{m^2}{2}\int\d^4x\sqrt{-g}\Big\{(\na\psi)^2-(\na\phi)^2-m^2\,\phi^2 \Big\}\,.
\ee
This expression shows that the action~(\ref{eqn:toyact}) does actually describe two scalar degrees of freedom, and that, for $m^2>0$, the massless mode has a ``wrong sign'' kinetic term, {\em i.e.}, is a ghost field. Now we solve the $\psi$- and $\phi$-equations of motion (EOMs):
\be
\Box\,\psi = 0\,,\quad (\Box-m^2)\,\phi=0\,,
\ee
or
\bea
\hat\psi'' - \frac{2}{\tau}\hat\psi' + k^2\hat\psi=0\,, \quad
\hat\phi'' - \frac{2}{\tau}\hat\phi' + \big(k^2+\frac{m^2}{H^2\,\tau^2}\big)\hat\phi=0\,, 
\eea
where $\;\hat{}\;$ implies a Fourier transform along the three spatial dimensions and ${}'\equiv\frac{\d}{\d\tau}$. The solutions are
\bea
&&\hat\psi=(-\tau)^{3/2}\big\{a_1\, H^{(1)}_{3/2}(-k\tau)+b_1\, H^{(2)}_{3/2}(-k\tau)\big\}\,, \\
&&\hat\phi=(-\tau)^{3/2}\big\{a_2\, H^{(1)}_{\nu/2}(-k\tau)+b_2\, H^{(2)}_{\nu/2}(-k\tau)\big\}\,,
\eea
with $\nu\equiv \sqrt{9-4\,m^2/H^2}$. We will use as mode functions the positive frequency modes at small scales ($|k\,\tau|\gg 1$), so that we set $b_1=b_2=0$. 

To obtain a unitary theory we need to project out the massless mode $\hat{\psi}$. To this end we observe that, since the behavior of the solutions as $\tau\to 0^-$ is
\bea
&&\hat\psi \to -i\sqrt{\frac{2}{\pi}} \frac{a_1}{k^{3/2}}\,, \\
&&\hat\phi \to -\frac{i}{\pi} (-\tau)^{(3-\nu)/2}\Big(\frac{2}{k}\Big)^{\nu/2}\Gamma\big(\frac{\nu}{2}\big)\, a_2 \,, 
\eea
imposing the Dirichlet boundary conditions, $\hat\psi|_{\tau=0}=0$ and $\hat\phi|_{\tau=0}=0$, forces $a_1=0$, while the $\phi$-mode survives as long as Re$(\nu)<3$, {\em i.e.}, $m^2>0$. In terms of the original field $\Psi$, this amounts to imposing the Dirichlet boundary condition $\lim_{\tau\to 0^-}\frac{1}{(-H\tau)^2}\,\Psi= 0$. Therefore, although we started with a higher derivative theory, by imposing appropriate boundary conditions and by choosing the overall sign of the action (\ref{eqn:toyact}) such that the massive field $\phi$ is normal, we can obtain a unitary theory of a single massive field. Once we have selected the linearized modes of the theory this way, the effect of sources or of self-interactions of the $\Psi$ field can be accounted for in a perturbative sense, and the theory is guaranteed to be ghost-free at tree level. Of course, one has to make sure that the sources for $\Psi$ allow consistent imposition of the boundary conditions.

\subsection{Massive gravity with boundary conditions: Linearized analysis}\label{sec:massivelin}

The most general action for an ``$R^2$"-HDG in 4D contains a cosmological constant, $R$, $R^2$ and $R_{\mu\nu}R^{\mu\nu}$ ($R_{\mu\nu\rho\lambda}R^{\mu\nu\rho\lambda}$ can always be traded for $R^2$ and $R_{\mu\nu}R^{\mu\nu}$, since in 4D the Gauss-Bonnet invariant is a total derivative) and can be shown~\cite{stelle2} to describe a massless tensor, a massive tensor and a massive scalar. We will assume a special relation between the coefficients of $R_{\mu\nu}R^{\mu\nu}$ and $R^2$, with which the scalar mode gets infinitely heavy and is removed from the theory.\footnote{In 3D, the same consideration would fix the ratio of the coefficients of $R_{\mu\nu}R^{\mu\nu}$ and $R^2$ to be $-3/8$, which is the choice leading to the New Massive Gravity of~\cite{Bergshoeff:2009hq}.} We will discuss in \S\ref{sec:conc} our expectations in the case where $R_{\mu\nu}R^{\mu\nu}$ and $R^2$ carry arbitrary coefficients.

Based on the considerations above, we choose the action of our system to be
\be\label{eqn:fullact}
S = \frac{M_P^2}{2}\int\d^4x\sqrt{-g}\Big[ \sigma\, R - \Lambda + \alpha\,\Big( R_{\mu\nu}R^{\mu\nu} - \frac{1}{3} R^2\Big) \Big] + {\rm boundary\;terms}\,,
\ee
where $\sigma=+1$ or $-1$, which will be determined by requiring the massive sector to be normal. We perform a perturbative expansion of (\ref{eqn:fullact}) on top of the de Sitter background
\be
\d s^2 = \frac{1}{(-H\tau)^2}\,(\eta_{\mu\nu} + h_{\mu\nu})\,\d x^\mu \d x^\nu\,.
\ee
At ${\cal O}(h)$, tadpole cancellation fixes $\Lambda$ to be 
\be\label{eqn:tadpolecan}
\Lambda=6\,\sigma\,H^2 \,.
\ee

In order to count the number of the physical DOFs, we decompose $h_{\mu\nu}$ into different helicity modes:
\bea\label{eqn:heldcmp}
&& h_{00} = -2\,\phi\,,\;\;
h_{0i} = h_{0i}^{\rm T} + \del_i B\,, \nn\\
&& h_{ij} = h_{ij}^{\rm TT} + \del_i\xi_j^{\rm T} + \del_j\xi_i^{\rm T} -2\,\delta_{ij}\,\psi + 2\,\del_i\del_j E \,,
\eea
where $\del_i h_{0i}^{\rm T}=0$, $\delta_{ij}h_{ij}^{\rm TT}=0$, $\del_i h_{ij}^{\rm TT}=0$ and $\del_i\xi_i^{\rm T}=0$. By construction, there is no mixing between different helicities at the quadratic level, and we can consider the helicity--2, --1 and --0 sectors separately.

The helicity--2 sector is simple:
\be\label{eqn:hel2act}
S^{(2)}_{\rm hel-2} = \frac{M_P^2}{2}\int\d^4x\,\Big[\, 
\frac{\alpha}{4}\,(h_{ij}^{\rm TT}{}'')^2
+ \frac{1}{4}\, h_{ij}^{\rm TT}{}' \,\Big(2\,\alpha\,\Delta + \frac{\sigma}{H^2\,\tau^2}\Big) h_{ij}^{\rm TT}{}'+ \frac{1}{4}\,\Delta h_{ij}^{\rm TT}\, \Big(\alpha\,\Delta + \frac{\sigma}{H^2\tau^2}\Big)\, h_{ij}^{\rm TT} \,\Big] \,, 
\ee 
with $\Delta=\partial_i^2$. Introducing the auxiliary field $f_{ij}$ to eliminate higher $\tau$-derivatives turns (\ref{eqn:hel2act}) into
\bea\label{eqn:hel2intm}
S^{(2)}_{\rm hel-2} &=& \frac{M_P^2}{2}\int\frac{\d^4x}{\tau^2}\,\Big[\, 
\frac{\sigma-2\,H^2\,\alpha}{4\,H^2}\big\{(h_{ij}^{\rm TT}{}')^2 + h_{ij}^{\rm TT}\,\Delta h_{ij}^{\rm TT}\big\} \nn\\
&&\hspace{60pt}+ \frac{\alpha}{\sqrt{2}}\big(f_{ij}'\, h_{ij}^{\rm TT}{}' + f_{ij}\,\Delta h_{ij}^{\rm TT}\big) - \frac{\alpha}{2\,\tau^2}\,f_{ij}^2 \,\Big] \,,
\eea
and diagonalizing it gives
\bea\label{eqn:hel2act2td}
&&S^{(2)}_{\rm hel-2} = \frac{M_P^2}{2} \frac{\mu^2\,(\sigma - 2\,H^2\,\alpha)}{2} \nn\\
&&\qquad\int\d^4x\,\Big[\, \frac{1}{2} (X_{ij}')^2 + \frac{1}{2}X_{ij}\, \Big(\Delta+\frac{2}{\tau^2}\Big) X_{ij} - \frac{1}{2} (Y_{ij}')^2 - \frac{1}{2}Y_{ij}\,\Big(\Delta+\frac{2-m^2/H^2}{\tau^2}\Big) Y_{ij} \,\Big]\,,\nn\\
\eea 
where $h_{ij}^{\rm TT} = -H\,\tau\cdot\mu\,(X_{ij} + Y_{ij})$, $f_{ij}=-H\,\tau\cdot\frac{\mu}{\sqrt{2}}\Big(2-\frac{\sigma}{H^2\alpha}\Big)\,Y_{ij}$, $m^2\equiv 2\,H^2-\frac{\sigma}{\alpha}$ and $\mu$ is an arbitrary normalization constant. To determine which modes are ghosts, we employ the Hamiltonian formulation with $S^{(2)}_{\rm hel-2} = \frac{M_P^2}{2} \frac{\mu^2}{2}\int\d^4x\,{\cal L}^{(2)}_{\rm hel-2}$:
\bea\label{eqn:hel2ham}
S^{(2)}_{\rm hel-2} &=& \frac{M_P^2}{2} \frac{\mu^2}{2} 
\int\d^4x\,\Big[\, P_X X'-\Big\{\frac{P_X^2}{2\,(\sigma - 2\,H^2\,\alpha)}-\frac{\sigma - 2\,H^2\,\alpha}{2}\,X\,\Big(\Delta+\frac{2}{\tau^2}\Big)X \Big\} \nn\\
&&\quad+P_Y Y' -\Big\{-\frac{P_Y^2}{2\,(\sigma - 2\,H^2\,\alpha)}+\frac{\sigma - 2\,H^2\,\alpha}{2}\,Y\,\Big(\Delta+\frac{2-m^2/H^2}{\tau^2}\Big)Y \Big\}\,\Big]\,,\qquad
\eea 
where $P_{X,Y}\equiv\frac{\delta{\cal L}^{(2)}_{\rm hel-2}}{\delta (X,Y)}$ and $ij$ indices are suppressed. Therefore in order for the Hamiltonian of the \emph{massive} helicity--2 modes to be positive definite, we have to require
\be\label{eqn:hel2pos}
\sigma - 2\,H^2\,\alpha < 0\,.
\ee
Notice that this implies the massless helicity--2 modes are ghost-like.

The helicity--1 sector, once the gauge invariant variable is identified, is even simpler. With $v_i \equiv \sqrt{-\Delta}\,(h_{0i}^{\rm T} - \xi_i^{\rm T}{}')$, we have
\bea\label{eqn:hel1act}
S^{(2)}_{\rm hel-1} &=& \frac{M_P^2}{2}\,\alpha \int\d^4x\,\Big[\, 
\frac{1}{2} (v_i')^2 + \frac{1}{2} v_i\,\Big(\Delta+\frac{2-m^2/H^2}{\tau^2}\Big)\, v_i \,\Big] \,. 
\eea
From the Hamiltonian analysis similar to that of the helicity--2 modes, $\alpha$ must satisfy
\be\label{eqn:hel1pos}
\alpha > 0\,.
\ee

The helicity--0 sector requires some work. Employing the gauge invariant variables
\be
\Phi = \phi - \tau\,\psi' \,,\quad
{\cal B} = \Delta(B - E' + \tau\,\psi) \,,
\ee
the helicity--0 part of (\ref{eqn:fullact}) is\footnote{As mentioned earlier, instead of (\ref{eqn:fullact}) if we had started with
\be
S = \frac{M_P^2}{2}\int\d^4x\sqrt{-g}\Big( \sigma R - \Lambda + \alpha R_{\mu\nu}R^{\mu\nu} - \beta R^2 \Big)\,,\nn
\ee
with $\alpha \neq 3\beta$, there would have been two propagating scalar DOFs. In the current formulation, this would have been manifested by the presence of a higher $\tau$-derivative term: $(\alpha-3\beta)(\psi'')^2$.}
\bea
S^{(2)}_{\rm hel-0} &=& \frac{M_P^2}{2} \int\d^4x\,\Big[\,
\frac{2}{3}\,\alpha\,({\cal B}')^2 + \frac{4}{3}\,\alpha\,{\cal B}'\,\Delta\Phi + \frac{4\,\sigma}{H^2\,\tau^3}\,{\cal B}\,\Phi + \frac{2}{3}\,\Phi\,\Big(\alpha\,\Delta^2-\frac{9\,\sigma}{H^2\,\tau^4}\Big)\Phi \,\Big].\nn\\
\eea
Note that $\Phi$ is non-dynamical. With $S^{(2)}_{\rm hel-0} \equiv \frac{M_P^2}{2} \int\d^4x\,{\cal L}^{(2)}_{\rm hel-0}$, writing in Hamiltonian formulation gives
\be
S^{(2)}_{\rm hel-0} = \frac{M_P^2}{2} \int\d^4x\,\Big[\,
P_B \,{\cal B}' - \frac{3}{8\,\alpha}\,P_B^2 + P_B\,\Delta\Phi + \frac{2\,\sigma}{H^2\,\tau^4}\,\Phi\,(2\,\tau\,{\cal B} - 3\,\Phi) \,\Big]\,.
\ee
Since $\Phi$ is a constraint, we integrate it out to get
\be
S^{(2)}_{\rm hel-0} = \frac{M_P^2}{2} \int\d^4x\,\Big[\,
P_B \,{\cal B}' - \Big(\frac{3}{8\,\alpha} - \frac{H^2\,\tau^4\,\Delta^2}{24\,\sigma}\Big)\,P_B^2 + \frac{\tau}{3}\,P_B\,\Delta{\cal B} + \frac{2\,\sigma}{3\,H^2\,\tau^2}{\cal B}^2 \,\Big]\,.
\ee
Upon a canonical transformation, 
\be
P_B = -\frac{2\,\sigma}{\sqrt{3}\,H\,\tau}\,Q \,, \quad
{\cal B} = \frac{\sqrt{3}\,H\,\tau}{2\,\sigma}P + \Big(\frac{\sqrt{3}\,H}{2} + \frac{H\,\tau^2\,\Delta}{2\,\sqrt{3}}\Big)Q \,,
\ee
we finally obtain
\be
S^{(2)}_{\rm hel-0} = \frac{M_P^2}{2} \int\d^4x\,\Big[\,
P\, Q' - \Big\{\frac{P^2}{2(-\sigma)}-\frac{(-\sigma)}{2}\,Q\,\Big(\Delta+\frac{2-m^2/H^2}{\tau^2}\Big)\,Q\Big\} \,\Big]\,,
\ee
which dictates $\sigma$ to be negative, \ie, 
\be\label{eqn:hel0pos}
\sigma=-1 \,.
\ee

Let us then combine the results from all three sectors. The positivity conditions for three sectors, (\ref{eqn:hel2pos}), (\ref{eqn:hel1pos}) and (\ref{eqn:hel0pos}), can be simultaneously satisfied, and we choose $\mu=\sqrt{\frac{2}{2\,H^2\alpha+1}}$, $F_2\equiv Y$, $F_1\equiv \sqrt{\alpha}\,v$ and $F_0\equiv Q$, to write the full quadratic action as
\be
S^{(2)} = \frac{M_P^2}{2}\int\d^4x\,\Big[\, 
\sum_{n=0}^2\Big\{\frac{1}{2} (F_n')^2 + \frac{1}{2} F_n\Big(\Delta+\frac{2-m^2/H^2}{\tau^2}\Big) F_n\Big\}
-\frac{1}{2}\dot X^2 - \frac{1}{2}X \Big(\Delta+\frac{2}{\tau^2}\Big) X \, \Big] \,,
\ee
with 
\be
m^2= 2\,H^2+\frac{1}{\alpha}\,.
\ee
That is, there are a normal massive spin--2 field with 5 DOFs and a ghost massless helicity--2 mode with 2 DOFs. Note that the Higuchi ghost~\cite{Higuchi:1986py} is not present because $m^2 > 2\,H^2$. 

The solutions of the equations of motion for the Fourier modes $\hat F_n$ and $\hat X$ are
\be
\hat F_n = \sqrt{-\tau}\,\big\{a_n\, H^{(1)}_{\nu/2}(-k\tau) + b_n \,H^{(2)}_{\nu/2}(-k\tau)\big\} \,, \quad
\hat X = \sqrt{-\tau}\,\big\{c\, H^{(1)}_{3/2}(-k\tau) + d\, H^{(2)}_{3/2}(-k\tau)\big\} \,,
\ee
with $\nu=\sqrt{9-4\,m^2/H^2}=\sqrt{1-4/(\alpha\,H^2)}$. Requiring the standard short scale behavior immediately sets $b_n=d=0$. 

As we did for the scalar toy model, we now use boundary conditions to eliminate the unwanted ghost.\footnote{See, \eg, \cite{Anninos:2011jp} for issues regarding imposing future boundary conditions in dS spacetime and their possible resolutions.} Since we impose boundary conditions on the metric perturbation $h_{\mu\nu}$ and the leading $\tau$-dependences for small $|\tau|$ are $F_2$, $X \sim \tau^{-1}h^{\rm TT}_{ij}$, $F_1 \sim h^{\rm T}_{0i}$ and $F_0 \sim B$, we need to know the behavior of $\tau F_2$, $\tau X$, $F_1$ and $F_0$ as $\tau\to0^-$:
\bea
-\tau\hat F_2 &\to& -\frac{i}{\pi} (-\tau)^{(3-\nu)/2}\Big(\frac{2}{k}\Big)^{\nu/2}\Gamma\big(\frac{\nu}{2}\big) a_2 \,,\quad \\
\hat F_1 &\to& -\frac{i}{\pi} (-\tau)^{(1-\nu)/2}\Big(\frac{2}{k}\Big)^{\nu/2}\Gamma\big(\frac{\nu}{2}\big) a_1 \,,\quad \\
\hat F_0 &\to& -\frac{i}{\pi} (-\tau)^{(1-\nu)/2}\Big(\frac{2}{k}\Big)^{\nu/2}\Gamma\big(\frac{\nu}{2}\big) a_0 \,,\quad \\
-\tau\hat X &\to& -i\sqrt{\frac{2}{\pi}}\frac{c}{k^{3/2}}\,.
\eea
Therefore we see that, if we impose the Dirichlet boundary condition on $h_{\mu\nu}$ at $\tau= 0^-$, the five massive modes can survive because ${\rm Re}(\nu)<1$, but the massless ones are projected out, and we are left with a unitary massive spin-2 system.

\section{Boulware-Deser ghost: Full ADM analysis}\label{sec:adm}%

In addition to the higher derivative ghost, HDG, like any theory of massive gravity, might be haunted by the BD ghost. Since the BD mode shows up at the nonlinear level, the best way to see if it is there is to employ the full Arnowitt-Deser-Misner (ADM) formulation~\cite{Arnowitt:1962hi} and count the number of Hamiltonian constraints.

With the introduction of an auxiliary metric, $f_{\mu\nu}$, (\ref{eqn:fullact}) can be written as
\be
S = \frac{M_P^2}{2}\int\d^4 x \sqrt{-g}\,\Big\{ \sigma R + f_{\mu\nu} \,\big(R^{\mu\nu} - \frac{1}{2}\,g^{\mu\nu}\,R\big) - \frac{1}{4\,\alpha}\,(f_{\mu\nu}\,f^{\mu\nu} - f^2) \Big\} \,.
\ee
For ADM treatment, we first decompose the metrics into ADM variables:
\be\label{eqn:admmetric}
g_{\mu\nu} = \begin{pmatrix}
-N^2+N_i\, N^i\; & N_i \\ N_j & \Gamma_{ij} \end{pmatrix} \,,\quad
f_{\mu\nu} = \begin{pmatrix}
-2\,N\,n + 2\,N_i\,n^i + \gamma_{ij}\,N^i\,N^j\; & n_i + \gamma_{ik}\,N^k \\ n_j + \gamma_{jk}\,N^k & \gamma_{ij} \end{pmatrix} \,,
\ee
to obtain
\bea\label{eqn:admact}
S &=& \frac{M_P^2}{2}\int\d^4 x\, N\,\sqrt{\Gamma}\, \Big\{ \Big(\sigma - \frac{n}{N}- \frac{\gamma}{2}\Big)\, {}^{(3)}R + \Delta\gamma - \nabla_i\nabla_j\gamma^{ij} + {}^{(3)}R_{ij}\,\gamma^{ij} \nn\\
&&\qquad - \frac{1}{4\,\alpha}\,\Big(\gamma_{ij}\,\gamma^{ij} - \gamma^2 - \frac{2\,n_i\,n^i}{N^2} - \frac{4\,\gamma\, n}{N}\Big) \nn\\
&&\qquad + \Big(\sigma + \frac{n}{N} - \frac{\gamma}{2}\Big)\,(K_{ij}\,K^{ij} - K^2) - 2\,K\, K_{ij}\,\gamma^{ij} + 2\,\gamma_{ij}\,K^i_k\,K^{jk} \nn\\
&&\qquad + 2\,K\,k - 2\,K_{ij}\,k^{ij} + \frac{2}{N}\,(K\,\nabla_in^i - K^{ij}\,\nabla_in_j) \Big\} \,,
\eea
where
\bea
K_{ij} &=& \frac{1}{2N}(-\dot\Gamma_{ij}+\nabla_iN_j+\nabla_jN_i) \,,\\
k_{ij} &=& \frac{1}{2N}(-\dot\gamma_{ij}+N^k\,\nabla_k\gamma_{ij}+\gamma_{ki}\,\nabla_jN^k+\gamma_{kj}\,\nabla_iN^k)\,.
\eea
Three-dimensional indices are raised and lowered by $\Gamma_{ij}$, so that $K\equiv \Gamma^{ij}\,K_{ij}$, $k\equiv \Gamma^{ij}\,k_{ij}$ and $\gamma\equiv \Gamma^{ij}\,\gamma_{ij}$. By defining the canonical momenta conjugate to $\Gamma_{ij}$ and $\gamma_{ij}$ by 
\be
P_{ij}\equiv\frac{\delta}{\delta\dot\Gamma_{ij}}\Big(\frac{2}{M_P^2}\,S\Big)\,,\quad 
\Pi_{ij}\equiv\frac{\delta}{\delta\dot\gamma_{ij}}\Big(\frac{2}{M_P^2}\,S\Big)\,,
\ee
we can put (\ref{eqn:admact}) into the canonical form 
\be
S = \frac{M_P^2}{2}\int\d^4x (P^{ij}\,\dot\Gamma_{ij} + \Pi^{ij}\,\dot\gamma_{ij} - {\cal H}) \,,
\ee
with
\bea
&&{\cal H} = -2\,n_i\,\nabla_jP^{ij} + N^k\,\big\{P^{ij}\,\nabla_k\gamma_{ij} - 2\,\nabla_j(P^{ij}\gamma_{ik}) - 2\,\nabla_i\Pi^i_k\big\} \nn\\
&&\quad - \sqrt{\Gamma}\,\Big\{\Big(\sigma - \frac{n}{N}- \frac{\gamma}{2}\Big){}^{(3)}R + \Delta\gamma - \nabla_i\nabla_j\gamma^{ij} + {}^{(3)}R_{ij}\,\gamma^{ij} - \frac{1}{4\,\alpha}\,\Big(\gamma_{ij}\,\gamma^{ij} - \gamma^2 - \frac{2\,n_i\,n^i}{N^2} - \frac{4\,\gamma \,n}{N}\Big)\Big\} \nn\\
&&\quad - \frac{N}{\sqrt{\Gamma}}\,\Big\{2\,P^{ij}\,\Pi_{ij} - P\,\Pi + \Big(\sigma + \frac{n}{N} - \frac{\gamma}{2}\Big)\,\Big(P^{ij}\,P_{ij} - \frac{P^2}{2}\Big) + 2\,P_{ij}\,P^{jk}\,\gamma^i_k - P\,P^{ij}\,\gamma_{ij}\Big\} \,. 
\eea
After integrating out the constraint $n_i$, we finally obtain 
\be
{\cal H} = N\,{\cal H}_0 + N_i\,{\cal H}_i + n\,\tilde{\cal H}_0\,,
\ee
where
\bea
{\cal H}_0 &=& \frac{1}{\sqrt{\Gamma}}\Big\{P\Pi - 2P^{ij}\Pi_{ij} + \Big(\sigma - \frac{\gamma}{2}\Big)\Big(\frac{P^2}{2} - P^{ij}P_{ij}\Big) + PP^{ij}\gamma_{ij} - 2P_{ij}P^{jk}\gamma^i_k + 2\alpha\nabla_iP^{ik}\nabla_jP^j_k\Big\} \nn\\
&& - \sqrt{\Gamma}\Big\{\Big(\sigma-\frac{\gamma}{2}\Big){}^{(3)}R + \Delta\gamma - \nabla_i\nabla_j\gamma^{ij} + {}^{(3)}R_{ij}\gamma^{ij} - \frac{1}{4\alpha}(\gamma_{ij}\gamma^{ij} - \gamma^2)\Big\} \,, \\
{\cal H}_i &=& P^{jk}\nabla_i\gamma_{jk} - 2\nabla_j(P^{jk}\gamma_{ik}) - 2\nabla_j\Pi^j_i \,,\\
\tilde{\cal H}_0 &=& \sqrt{\Gamma}\Big({}^{(3)}R-\frac{1}{\alpha}\gamma\Big) - \frac{1}{\sqrt{\Gamma}}\Big(P^{ij}P_{ij}-\frac{P^2}{2}\Big) \,.
\eea
The lapse, $N$, and the shift, $N_i$, for $g_{\mu\nu}$ are Lagrange multipliers, as they should be: the resulting constraints immediately remove four DOFs, and additional four will be eliminated with secondary constraints. What is important to note is that the lapse, $n$, for $f_{\mu\nu}$ is also a Lagrange multiplier, so that two more DOFs will be killed by the corresponding primary and secondary constraints. Therefore, DOF counting goes as
\bea
&&6[\Gamma_{ij}] + 6[P_{ij}] + 6[\gamma_{ij}] + 6[\Pi_{ij}] - 2\times(1+3+1)[N,N_i,n]\nn\\
&&=(5+5)[{\rm massive\;\; graviton}] + (2+2)[{\rm massless\;\; graviton}]\,,\nn
\eea
showing that, even at the full nonlinear level, the action~(\ref{eqn:fullact}) describes 5+2 propagating degrees of freedom and there is no 6th massive mode, \ie, the BD ghost. Then imposing boundary conditions will remove the massless modes in the end.

\section{Discussion}\label{sec:conc}%

In this paper we have described a way of realizing a non-linear completion of PF massive gravity without ghosts. Yet, there are more items in the to-be-resolved list presented in the introduction. Let us briefly discuss a couple of them.

The vDVZ discontinuity is not supposed to be an issue in the present set-up, because the discontinuity disappears on (A)dS backgrounds as $m^2/H^2\to 0$~\cite{Kogan:2000uy}. However, in the case of dS background we have considered here the range $0<m^2<2\,H^2$ is forbidden{, since massive graviton would be a ghost in this regime; the conditions~(\ref{eqn:hel0pos}) $\sigma=-1$ and~(\ref{eqn:hel1pos}) $\alpha>0$ were imposed precisely to avoid the appearance of such a ghost. Therefore, the limit $m\to 0$ does not belong to the allowed parameter space for our scenario.

It is worth noting that HDG (``curvature squared gravity", to be precise) shares a set of solutions with GR. In  particular, in the presence of a spherically symmetric source, the Schwarzschild-dS (SdS) metric solves the EOMs of HDG and satisfies the Dirichlet boundary condition on dS boundary, which seems enough to make SdS a viable solution for our system. But it turns out not to be the case. In fact, our EOMs can be schematically written as
\be\label{eqn:scheom}
-G_{\mu\nu} + \frac{\Lambda}{2}\,g_{\mu\nu} + \alpha\, {\cal E}_{\mu\nu} = \frac{T_{\mu\nu}}{M_P^2}\,,
\ee
where $G_{\mu\nu}$ is the Einstein tensor, $\cal E_{\mu\nu}$ the contribution from the curvature-squared terms and $T_{\mu\nu}$ the stress energy tensor for a spherically symmetric source. For solutions of pure GR, $\cal E_{\mu\nu}$ identically vanishes, and we are left with an Einstein-like equation with an overall minus sign on the left hand side, which results in a \emph{negative} Schwarzschild radius. If we do not allow for naked singularities in the theory, SdS is not a valid solution of our system. It would be interesting to find a physical, asymptotically dS, spherically symmetric solution to (\ref{eqn:scheom}) and  to analyze its behavior at various scales.

As for the stability/robustness of the model, we stress that we are considering it only at the classical level (this is not more restrictive than what happens in the other massive gravity theories discussed in the introduction). In a general quantum mechanical theory, indeed, we would expect $R^3$, $R^4$, $\cdots$ corrections to be generated and to affect with ${\cal O}(1)$ corrections the properties of the massive mode. A related question concerns the robustness of the theory with respect to the tuning of the ratio, $-\frac{1}{3}$, of the coefficients of $R_{\mu\nu}R^{\mu\nu}$ and $R^2$ in the lagrangian, \ie, whether we can still have a unitary theory even if this tuning is broken. When we start with $\frac{M_P^2}{2}\int\d^4x\sqrt{-g}\big( \sigma\,R - \Lambda + \alpha\, R_{\mu\nu}\,R^{\mu\nu} -\beta \,R^2 \big)$, the theory contains an ``eight mode'' with mass proportional to $\left(\alpha-3\,\beta\right)^{-1}$~\cite{stelle2} that, for our choice $\sigma=-1$, should behave as a ghost. Now, in analogy with the analysis of section \ref{sec:massivelin}, one may speculate that the scalar ghost could be truncated by the Dirichlet boundary conditions -- at least in part of the parameter space spanned by $\alpha$ and $\beta$. It might be worthwhile to confirm this speculation with a full, explicit investigation.

\acknowledgments
We would like to thank Nemanja Kaloper and Ian-woo Kim for interesting discussions. L.S. thanks for hospitality the University of Paris-Sud, where part of this work was realized. This work is partially supported by the U.S. National Science Foundation grant PHY-1205986.

\end{document}